# Photonic Astronomy and Quantum Optics

Dainis Dravins

*Lund Observatory,  Box 43,  SE-22100 Lund, Sweden*

**Abstract.**  Quantum optics potentially offers an information channel from the Universe beyond the established ones of imaging and spectroscopy.  All existing cameras and all spectrometers measure aspects of the first-order spatial and/or temporal coherence of light.  However, light has additional degrees of freedom, manifest in the statistics of photon arrival times, or in the amount of photon orbital angular momentum.  Such quantum-optical measures may carry information on how the light was created at the source, and whether it reached the observer directly or via some intermediate process.  Astronomical quantum optics may help to clarify emission processes in natural laser sources and in the environments of compact objects, while high-speed photon-counting with digital signal handling enables multi-element and long-baseline versions of the intensity interferometer.  Time resolutions of nanoseconds are required, as are large photon fluxes, making photonic astronomy very timely in an era of large telescopes.

## 1  What Is Observed in Astronomy?

Almost all of astronomy depends on the interpretation of properties of electromagnetic radiation ("light") from celestial sources. The sole exceptions are neutrino detections; gravitational-wave searches; analyses of cosmic rays, meteorites, and other extraterrestrial materials; and in-situ studies of solar-system bodies. All our other understanding of the Universe rests upon observing and interpreting more or less subtle properties in the light reaching us from celestial bodies.

Astronomical telescopes are equipped with myriads of auxiliary instruments which, on first sight, may give an impression of being vastly different. However, a closer examination of the underlying physical principles reveals that they all are measuring either the spatial or temporal [first-order] coherence of light (or perhaps some combination of these).  All imaging devices (cameras, interferometers) are studying aspects of the spatial coherence (in various directions, and for different angular extents on the sky).  All spectrally analyzing devices measure aspects of the temporal coherence (with different temporal/spectral resolution, and in the different polarizations). Although a gamma-ray satellite may superficially look different from a long-baseline radio interferometer, the basic physical property they are measuring is the same.

For centuries, optical astronomy has developed along with optical physics: Galileo's telescope and Fraunhofer's spectroscope were immediately applied to astronomical problems.  However, the frontiers of $21^{st}$ century optical physics have now moved toward photonics and quantum optics, studying individual photons, and photon streams.  Those can be complex, carrying information beyond imaging, spectroscopy, or polarimetry. Different physical processes in the generation of light may cause quantum-statistical differences (e.g., different degrees of photon bunching in time) between light with otherwise identical spectrum, polarization, intensity, etc., and studies of such non-classical properties of light are actively pursued in laboratory optics.

Since almost all astronomy is based upon the interpretation of subtleties in the light from astronomical sources, quantum optics appears to have the potential of becoming another information channel from the Universe, fundamentally different from imaging and spectroscopy.  What astronomy could then be possible with quantum optics?

## 2 What Is *Not* Observed in Astronomy?

Figure 1 illustrates a type of problem often encountered in astrophysics: Light from various sources has been created through different (but typically unknown) physical processes: thermal radiation, stimulated emission, synchrotron radiation, etc. Now, assume one is observing these sources through "filters", adjusted so that all sources have the same size and shape, same intensity, same spectrum, and the same polarization. How can one then tell the difference when observing the sources from a great distance?

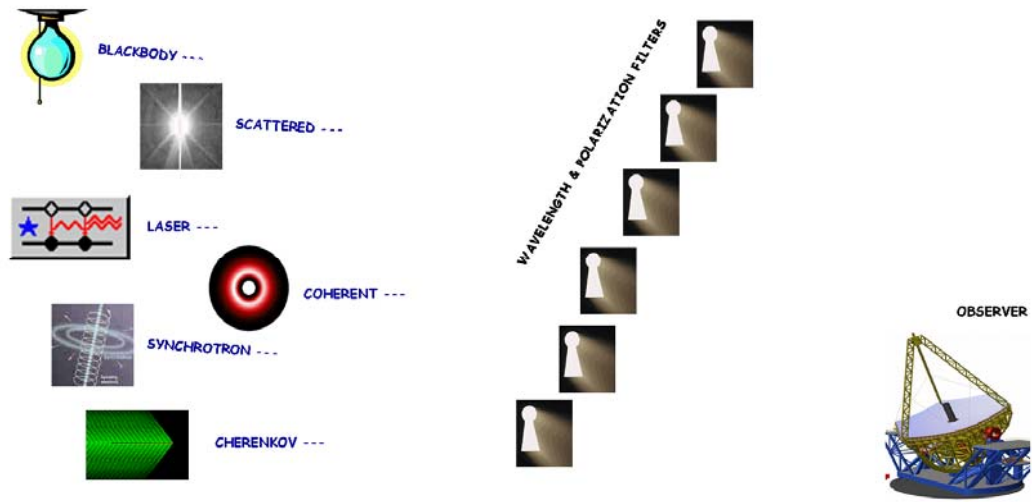

**Fig.1.** An observational problem beyond current imaging and spectroscopy: Various sources emit light created through different processes. If spatial and spectral filters give to all sources the same apparent size and shape, same intensity, same spectrum, and the same polarization, how can one then tell the difference?

For sources as defined, it actually is *not possible, not even in principle,* to segregate them using *any* classical astronomical instrument. Telescopes with imaging devices (cameras or interferometers) would record the same spatial image (of the keyhole aperture), and any spectrometer would find the same spectrum. Still, the light from those sources can be physically different since photons have more degrees of freedom than those relevant for mere imaging by telescopes or for spectroscopic analysis.

Identical images are produced by photons arriving from identical directions, and identical spectra by similar distributions of photon energies. However, a photon stream has further degrees of freedom, such as the *temporal statistics of photon arrival times*, giving a measure of ordering (entropy) within the photon-stream, and its possible deviations from "randomness". Such properties are reflected in the second- (and higher-) order coherence of light, observable as correlations between pairs (or a greater number) of photons, illustrated by a few examples in Fig. 2.

Clearly, the differences lie in *collective properties* of groups of photons, and cannot be ascribed to any one individual photon. The information content lies in the correlation in time (or space) between successive photons in the arriving photon stream (or the volume of a "photon gas"), and may be significant if the photon emission process has involved more than one photon.

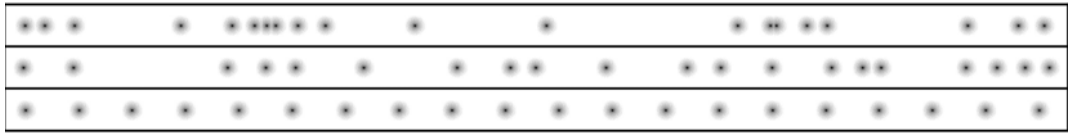

PHOTON ARRIVAL TIME

**Fig.2.** Statistics of photon arrival times in light beams with different entropies (different degrees of "ordering"). The statistics can be "quantum-random", as in maximum-entropy black-body radiation (following a Bose-Einstein distribution with a characteristic "bunching" in time; top), or may be quite different if the radiation deviates from thermodynamic equilibrium, e.g. for anti-bunched photons (where photons tend to avoid one another; center), or a uniform photon density as in stimulated emission from an idealized laser (bottom). The characteristic fluctuation timescales are those of the ordinary first-order coherence time of light (on order of only picoseconds for a 1 nm passband of optical light, but traces of which are detectable also with slower time resolutions). Figure in part adapted from Loudon [98]

The integer quantum spin of photons ($S = 1$) make them bosons, and the density and arrival-time statistics of a photon gas in a maximum-entropy ("chaotic")[1] state of thermodynamic equilibrium then obey a Bose-Einstein distribution, analogous to the Maxwell-Boltzmann distribution for classical particles in thermodynamic equilibrium. The statistics implies a certain bunching of photons, i.e., an enhanced probability of successive photons to arrive in pairs. This is probably the best-known non-classical property of light, first measured by Hanbury Brown and Twiss in those experiments that led to the astronomical intensity interferometer [55]. However, a photon gas that is in a different entropy state does not have to obey that particular distribution, just like classical particles do not necessarily have to obey a Maxwell-Boltzmann distribution.

A well-known example is that of stimulated emission from a laser: in the ideal case its photons display no bunching whatsoever, rather arriving in a uniform stream. This is a non-chaotic state of light, far from a maximum-entropy condition. For such emission to occur, there must be at least two photons involved: one that is the stimulating one, and another that has been stimulated to become emitted. The emerging light then contains (at least) two photons that are causally connected and mutually correlated. One single photon cannot alone carry the property of stimulated emission: that requires multiple photons for its description. On a fundamental level, the phenomenon is enabled by the boson nature of photons which permits multiple particles to occupy the same quantum state. Another related example is superfluidity in liquid helium (of the isotope $He^4$). A volume of superfluid helium-4 obtains very special properties; however a single helium atom cannot alone carry the property of "superfluidity"; that is a *collective* property of many helium atoms, revealed by their mutual correlations. The state of superfluidity in a volume of liquid helium-4 is quite analogous to the state of stimulated emission in a volume of photon gas; both are Bose-Einstein condensations, where the respective boson fluid has a uniform particle density throughout. A practical difference is that, while the liquid helium is stationary inside its container[2], the photon gas if flowing past any detector at the speed of light, and requires detectors with very rapid response for its density fluctuations to be revealed.

These quantum fluctuations in light are fully developed over timescales equal to the inverse bandwidth of light. For example, the use of a 1 nm bandpass optical filter gives a frequency bandwidth of $10^{12}$ Hz, and the effects are then fully developed on timescales of $10^{-12}$ seconds. While instrumentation with continuous data processing facilities with such resolutions is not yet available, it is possible to detect the effects, albeit with a decreased amplitude, also at the more manageable nanosecond timescales.

---

[1] In quantum optics, the term 'chaotic' denotes light from randomly fluctuating emitters, not to be confused with dynamically 'chaotic' states in mechanics
[2] except, perhaps, climbing its walls!

By contrast, particles with half-integer quantum spin (e.g., electrons with $S = ½$) are fermions, and obey different quantum statistics. In particular, groups of electrons suffer from the Pauli exclusion principle, which prohibits them to occupy the same quantum state. Thermal electrons therefore tend to avoid one another, analogous to the center illustration in Fig. 2. Photons enjoy much more freedom, and can appear either as preferentially bunched together in space and time, anti-bunched, or almost any other distribution.

Various processes studied in the laboratory, such as [multiple] scattering (in astrophysical jargon: "frequency redistribution"), passage through beamsplitters (in astrophysical jargon: "angular redistribution") modifies the relative amount of bunching, in principle carrying information of the events a photon stream has experienced since its creation (e.g., [3]). Likewise, light created under special conditions (free-electron laser, resonance fluorescence, etc.; in astrophysical jargon: "non-LTE") may have their own characteristic photon statistics, as discussed below. Thus (at least in principle), quantum statistics of photons may permit to segregate circumstances such as whether the Doppler broadening of a spectral line has been caused by motions of those atoms that emitted the light, or by those intervening atoms that have scattered the already existing photons.

## 3  The Intensity Interferometer

Although no existing astronomical instrument is capable of directly segregating the sources in Fig. 1, there actually has been one which in principle has had that capability: the stellar *intensity interferometer*, developed years ago in Australia by Hanbury Brown, Twiss, et al., for the original purpose of measuring stellar sizes [55]. Today this would be considered a quantum-optical instrument. At the time of its design, the understanding of its functioning was the source of considerable confusion, with numerous published papers questioning its basic principles. Indeed, at that time (and even now!), the intensity interferometer has been an instrument whose functioning is challenging to intuitively comprehend.[3]

To begin with, the name itself is sort of a misnomer: there actually is nothing interfering in the instrument; rather its name was chosen for its analogy to the ordinary [phase] interferometer, whose scientific aims the original intensity interferometer was replicating (in measuring stellar diameters). Two telescopes are simultaneously measuring the random and very rapid quantum fluctuations in the light from some particular star. When the telescopes are placed next to one another, both measure the same signal, but when moving them apart, the fluctuations gradually become decorrelated; how rapidly this occurs as the telescopes are moved apart gives a measure of the size of the star. The spatial baselines are the same as would be needed in ordinary phase interferometry, but the signal observed is thus the *correlation* between the fluctuations electronically measured in each of the two telescopes, and how this correlation gradually decreases as the telescopes are moved apart from one another.

One reason for the incomprehensibility was that, at the time the optical version of the instrument was developed (following earlier radio experiments), the quantum optical properties of light were still incompletely known, and novel light sources such as the laser were not yet fully developed. Especially confusing was to comprehend how *two* telescopes could simultaneously observe the "same" light from a star. Given that light consisted of photons, how could then the same photon be simultaneously detected by two different telescopes?

---

[3] The author actually has heard the late Robert Hanbury Brown himself make the remark that *"this must be the most incomprehensible instrument in all of astronomy"*.

It was to be some years until a good explanation and theory could be developed. Not only did this describe the functioning of that instrument, but it also led to a more complete quantum mechanical description of the nature of light. In particular, the resulting quantum theory of optical coherence shows that an arbitrary state of light can be specified with a series of coherence functions essentially describing one-, two-, three-, etc. -photon-correlations. While the ordinary (first-order) coherence can be manifest as the interference in amplitude and phase between light waves, the second-order coherence is manifest as either the correlation between photon arrival rates at different locations at any one given time (second-order *spatial* coherence, the effect exploited in the intensity interferometer), or as correlations in photon arrival rates between different delay times along the light beam, at any one given spatial location. The latter gives the second-order *temporal* coherence, an effect exploited in photon-correlation spectroscopy for measuring the wavelength width of scattered light. A key person for the development of the quantum theory of optical coherence was Roy Glauber [47–51], whose pioneering efforts were rewarded with the 2005 Nobel Prize in physics.

The functioning of the intensity interferometer is now well documented, and will not be repeated here. For details of the original instrument, see the original papers by Hanbury Brown et al., as well as retrospective overviews [55–62]. The principles are explained in various textbooks and reference publications, e.g., [52, 98, 102], and very lucidly in Labeyrie et al [83].

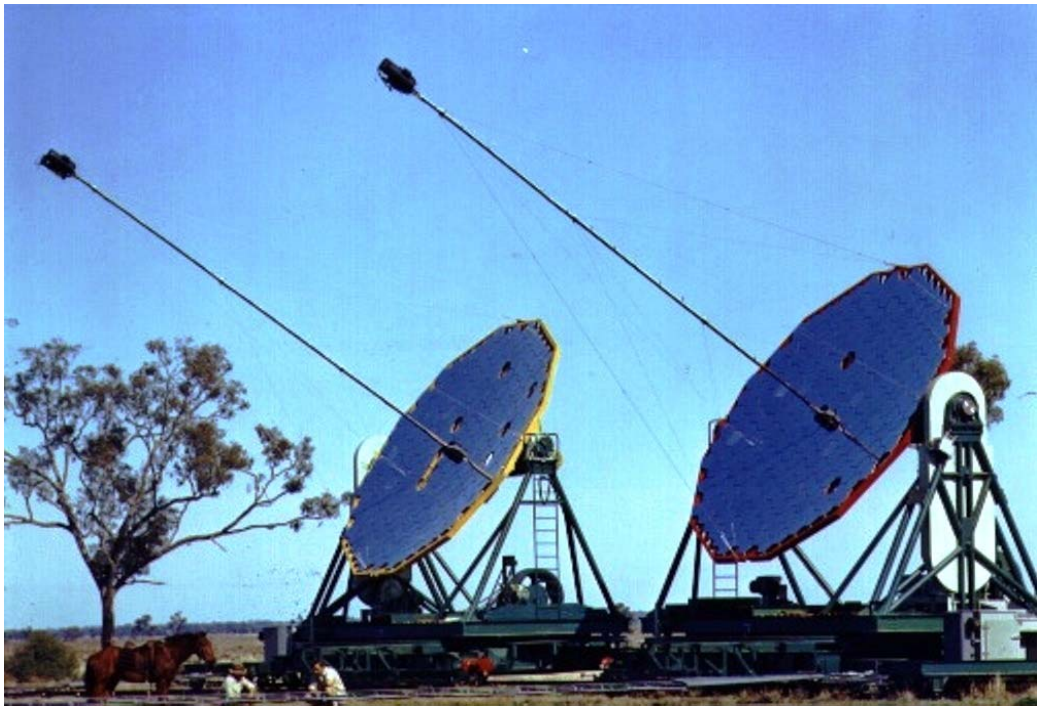

**Fig.3.** The only previous astronomical instrument which could have segregated among the light sources of Fig. 1. The original *intensity interferometer* at Narrabri, Australia, observed one star simultaneously with two telescopes, whose mutual distance was gradually changed. (University of Sydney; reprinted with permission from its School of Physics)

The significant observational advantage is that telescopic optical quality and control of atmospheric path-lengths need only to correspond to some [reasonable] fraction of the electronic resolution. For a realistic value of 10 ns, say, the corresponding light-travel distance is 3 m, and optical errors of maybe one tenth of this can be accepted, enabling coarse flux collectors to be used (rather than precise telescopes), and avoiding any sensitivity to atmospheric seeing (thus enabling very long baselines). The price to be paid is that large photon fluxes are required (leading to bright limiting magnitudes) and that (for a two-telescope system) only the modulus of the visibility function is obtained, giving a measure of the extension of the source, but not its full phase-correct image.

The quantity obtained from an intensity interferometer is the normalized *second-order correlation function* of light with its time-variable intensity $I(t)$:

$$g^{(2)}(\tau) = \frac{\langle I(t)I(t+\tau)\rangle}{\langle I(t)\rangle^2} \; ,$$

where $\tau$ is the correlation time-delay, $t$ is time, and $\langle \; \rangle$ denotes long-term averaging.

Although its function was perhaps not fully appreciated at the time it was developed, an intensity interferometer works correctly only for sources whose light is in the maximum entropy, thermodynamic equilibrium state. For such light, simple relations exist between the modulus of the first-order coherence function (the visibility measured in ordinary phase interferometers) and higher-order correlation functions. The relation between the normalized second-order correlation $g^{(2)}$ and the ordinary first-order quantity is $g^{(2)} = |g^{(1)}|^2$, so that the modulus of the visibility $g^{(1)}$ can be deduced, yielding stellar diameters from intensity correlation measurements.[4]

For a stable wave (such as an idealized laser), $g^{(2)}(\tau) = g^{(2)}(0) = 1$ for all time delays; while for chaotic, maximum-entropy light $g^{(2)}(0) = 2$, a value reflecting the degree of photon bunching in the Bose-Einstein statistics of thermodynamic equilibrium. In the laboratory, one can follow how the physical nature of the photon gas gradually changes from chaotic ($g^{(2)} = 2$) to ordered ($g^{(2)} = 1$) when a laser is "turned on" and the emission gradually changes from spontaneous to stimulated (although the spectrum does not change). Measuring $g^{(2)}$ and knowing the laser parameters involved, it is possible to deduce the atomic energy-level populations, which is an example of a parameter of significance to theoretical astrophysics ("non-LTE departure coefficient") which cannot be directly observed with any classical measurements of one-photon properties. Chaotic light that has been scattered against a Gaussian frequency-redistributing medium obtains a higher degree of photon bunching: $g^{(2)}(0) = 4$; while fully antibunched light has $g^{(2)}(0) = 0$. The latter state implies that whenever there is one photon at some time $= t$ [then $I(t) = 1$], there is none immediately afterwards, i.e., $I(t+\tau) = 0$ for sufficiently small $\tau$. Experimentally, this can be produced in, e.g., resonance fluorescence, and is seen through sub-Poissonian statistics of recorded photon counts, i.e. narrower distributions than would be expected in a "random" situation. Corresponding relations exist for higher-order correlations, measuring the properties for groups of three, four, or a greater number of photons. For introductions to the theory of such quantum-optical phenomena, see, e.g., [42, 97, 98, 102, 105, 121]. Experimental procedures are described by Bachor & Ralph [3], Becker [6], and Saleh [132].

The functioning of the intensity interferometer implicitly assumes the photon statistics of starlight to correspond to the Bose-Einstein distribution characteristic for thermodynamic equilibrium, as from a black-body radiator (top plot in Fig. 2). A hypothetical star shining with idealized laser light (as bottom plot in Fig. 2) would produce no intensity fluctuations whatsoever: no matter over what baseline one would place the telescopes, the instrument would not yield any meaningful signal from which to deduce the angular size of the star. The intensity interferometer was not designed for such purposes, but it could in principle have been used to search for possible deviations from randomness in the photon streams. For example, if the stellar diameter is determined independently (using ordinary phase interferometry, say), the difference between that value, and that deduced from the second-order coherence could be interpreted as a due to a difference in photon statistics and in the entropy of light.

---

[4] The expressions given here are simplified ones; a full quantum-optical description involves photon annihilation operators, etc.

## 4  Quantum Phenomena in Astronomy

While the existence *in principle* of various quantum-optical phenomena cannot be avoided, it is less obvious which astronomical objects will in practice be sources of measurable amounts of non-trivial photon-stream properties. In this Section, we review various studies of radiation processes where such phenomena could be expected.

What is the quantum nature of the light emitted from a volume with departures from thermodynamic equilibrium of the atomic energy level populations? Will a spontaneously emitted photon stimulate others, so that the path where the photon train has passed becomes temporarily deexcited and remains so for perhaps a microsecond until collisions and other effects have restored the balance? Does then light in a spectral line perhaps consist of short photon showers with one spontaneously emitted photon leading a trail of others emitted by stimulated emission? Such amplified spontaneous emission ("partial laser action") might occur in atomic emission lines from extended stellar envelopes or stellar active regions. Predicted locations are mass-losing high-temperature stars, where the rapidly recombining plasma in the stellar envelope can act as an amplifying medium [84, 158, 159]. Analogous effects could exist in accretion disks [38]. In the infrared, several cases of laser action are predicted for specific atomic lines [39, 54, 117]. Somewhat analogous situations (corresponding to a laser below threshold) have been studied in the laboratory. The radiation from "free" clouds (i.e. without any laser resonance cavity) of excited gas with population inversion is then analyzed. One natural mode of radiative deexcitation indeed appears to be the emission of "photon showers" triggered by one spontaneously emitted photon which is stimulating others along its flight vector out from the volume (Fig. 4).

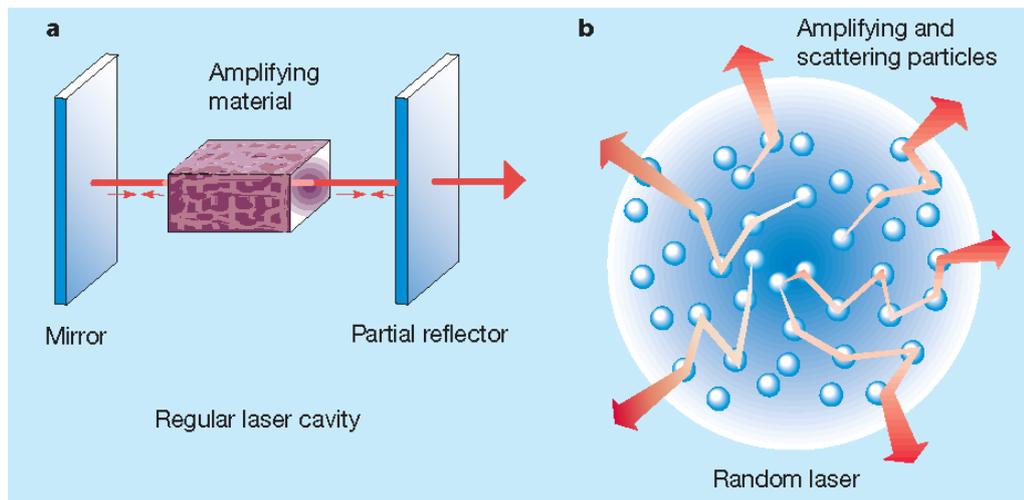

**Fig.4.** Different geometries producing laser emission in the laboratory, and in space ("random laser"): Wiersma [163]. Reprinted with permission from Macmillan Publishers Ltd: Nature **406**, 132, © 2000

Geometric differences between a laboratory laser, and an astrophysical one include that the laboratory one is normally enclosed inside an elongated cavity, whose end mirrors have the purpose of both increasing the path-length traversed by light inside the amplifying medium, and serving as a wavelength filter. Cosmic lasers involve the same basic processes, but of course do not have mirrors in resonance cavities, rather relying on sheer path-length to achieve amplification. An astrophysical laser must be of a more "random" nature, presumably producing monochromatic flashes which are initiated by "lucky" photons inside the amplifying volume. The beams are emitted in narrow angles, and we only observe those beams that happen to be directed at the Earth; the vast majority which do not point towards us, remain invisible.

Following an early theoretical prediction [93], and an experimental realization in the laboratory [86] random lasers are currently receiving significant attention, not least in trying to understand their statistical properties and intrinsic intensity fluctuations [16, 40, 90, 106, 156].

At first sight, it might appear that light from a star should be nearly chaotic because of the very large number of independent radiation sources in the stellar atmosphere, which would randomize the photon statistics. However, since the time constants involved in the maintenance of atomic energy level overpopulations (e.g. by collisions) may be longer than those of their depopulation by stimulated emission (speed of light), there may exist, in a given solid angle, only a limited number of radiation modes reaching the observer in a given time interval (each microsecond, say) and the resulting photon statistics might then well be non-chaotic. Proposed mechanisms for pulsar emission include stimulated synchrotron and curvature radiation ("free-electron laser") with suggested timescales of nanoseconds, over which the quantum statistics of light again would be non-chaotic. Obviously, the list of potential astrophysical targets can be made longer [27]. In general, photon statistics for the radiation from any kind of energetic source could convey something about the processes where the radiation was liberated.

## 4.1 Mechanisms producing astrophysical lasers

The possibility of *laser action*, i.e., an enhanced fraction of stimulated emission in certain spectral lines due to *amplified spontaneous emission*, has been suggested for a number of spectral lines in different sources. Already long ago it was realized that deviations from thermodynamic equilibrium in atomic energy-level populations could lead to such emission, although the possibility was not taken seriously before the construction of the laboratory microwave maser in 1954, and the visible-light laser in 1960. However, after the discovery in 1965 of the first celestial OH maser at λ 18 cm, the astrophysical possibilities become apparent [126].

Laser action normally requires a population inversion in atomic or molecular energy levels. Such deviations from thermodynamic equilibrium can be achieved through selective radiative excitation, or electronic recombination in a cooling plasma. Selective excitation, where another (often ultraviolet) emission line of a closely coinciding wavelength excites a particular transition, overpopulating its upper energy level, is often referred to as *Bowen fluorescence*. Following Bowen [12], plausible combinations of atomic and molecular lines have been studied by several, e.g., [13, 43].

Already long ago, a few authors did touch upon the possibility of laser action, but dismissed it as unrealistic or only a mathematical curiosity. Thus, Menzel [103], when discussing situations outside of thermodynamic equilibrium, concluded "*...the condition may conceivably arise when the value of the integral* [of absorbed radiation] *turns out to be negative. The physical significance of such a result is that energy is emitted rather than absorbed ... as if the atmosphere had a negative opacity. This extreme will probably never occur in practice.*" (ApJ **85**, p 335, 1937). However, 33 years later the same Donald Menzel [104] was a pioneer in proposing laser action in non-LTE atmospheres.

If a strongly ionized plasma is rapidly cooled, population inversion occurs during the subsequent electronic recombination cascade, permitting laser action. This scheme, known as plasma laser or recombination laser, has been studied in the laboratory (e.g., supersonic expansion of helium plasma from nozzles into low-density media), and is applicable to stars with mass loss. If the electron temperature is sufficiently low, the recombination occurs preferentially into the upper excited states of the ion, which then decay by a radiative and/or collisional cascade to the ground state. Within the cascade, population inversions may form among the excited states, depending upon the relative transition probabilities. Also, stepwise ionization implies the moving of electrons to higher energy levels, likewise tending to invert the energy-level populations.

Another mechanism for creating non-equilibrium populations of energy levels is an external X-ray illumination of stellar atmospheres, originating from a hot component in a close binary system [129].

Possibly, the earliest modeling of an astrophysical source in terms of laser action, was for the λ 190.9 nm C III intercombination line in the Wolf-Rayet star γ Vel by West in 1968 [162], although the interpretation was later doubted [19]. Several infrared atomic transitions with astrophysically favorable parameters for population inversion were identified by Smith [140], while Menzel [104] suggested laser action in non-LTE atmospheres by analyzing the microscopic form of the radiative transfer equation. Jefferies [70] showed that external pumping may not be required for close-lying atomic energy levels with energy separations $\Delta E$ much smaller than $kT$, with $T$ the kinetic temperature of the gas. Lavrinovich & Letokhov [84] suggested population inversions for lines such as O I 844.6 nm in the atmospheres of hot Be stars, while Peng & Pradhan [117] modeled infrared lasers in active galactic nuclei and novae. Selective excitation appears to cause the bright emission of ultraviolet Fe II lines in gas ejecta close to the central star of η Car [77] as well as in the symbiotic star RR Tel [64]. Masing in the forbidden [Fe XI] 6.08 μm line may be common, caused by peculiarities in the atomic energy structure [39]. Other authors have discussed infrared lasers in planetary atmospheres, active galactic nuclei, accretion systems and stellar envelopes; a review is by Johansson and Letokhov [76].

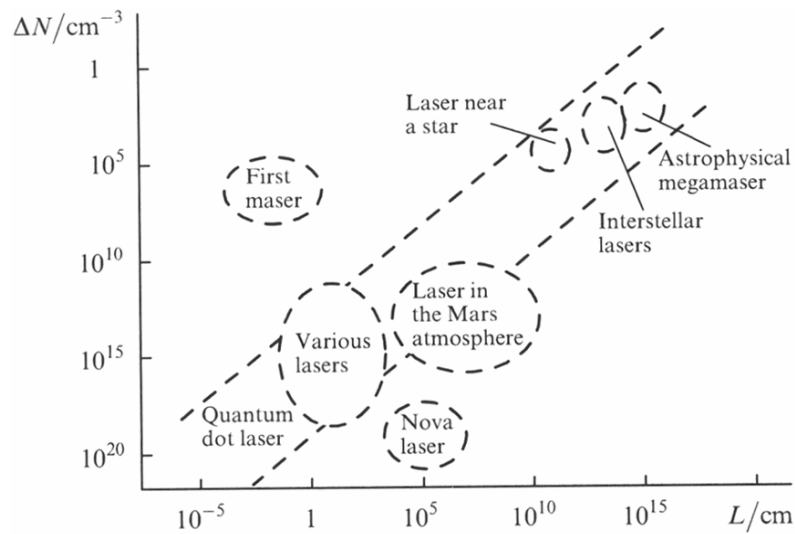

**Fig.5.** Various laboratory and astrophysical masers and lasers span very many orders of magnitude in the active-medium particle density – active-medium size diagram. $\Delta N$ denotes the density of atomic population inversions: Letokhov [94]. Reprinted from Quant Electr **32**, 1065 (2002), with permission from Turpion Ltd

For longer wavelengths, there is a broad literature on millimeter and radio masers, especially molecular ones. Millimeter recombination lines of high levels in hydrogen (H30 α, H50 β, and others) are seen towards η Car, apparently affected by maser emission [22], as well as in other sources.

Theoretical treatments of astrophysical radiative transfer including laser action have been made by several authors: [24, 38, 150, 151, 158, 159]. A common conclusion seems to be that while laser action seems possible under many diverse conditions, the amplification of any specific line requires a sufficiently extended region within a rather narrow range of parameter space (pressure, temperature, electron density).

Generally, radio and microwave masers can be produced more easily (gyrosynchrotron emission, curvature radiation) than those at shorter wavelengths [148], although there are suggestions for astrophysical lasers even in the X-ray regime, perhaps a recombination laser in hydrogen-like ions around accreting neutron stars [14, 44].

In the absence of feedback, the amplifying medium of an astrophysical laser represents an amplifier of spontaneous emission. Because of the large size of astrophysical lasers, the radiation pattern is determined by the geometry of the amplifying region. In the case of a nearly spherical geometry, the radiation should be isotropic.

## 4.2 Radiation from luminous stars

Eta Carinae, the most luminous star known in our Galaxy, is some 50–100 times more massive than our Sun and 5 million times as luminous ($M_{bol}$ approx= −12). This star is highly unstable, undergoing giant outbursts from time to time; one in 1841 created the bipolar Homunculus Nebula. At that time, and despite the comparatively large distance (3 kpc), η Car briefly became the second brightest star in the night sky. It is now surrounded by nebulosity expanding at around 650 km s$^{-1}$.

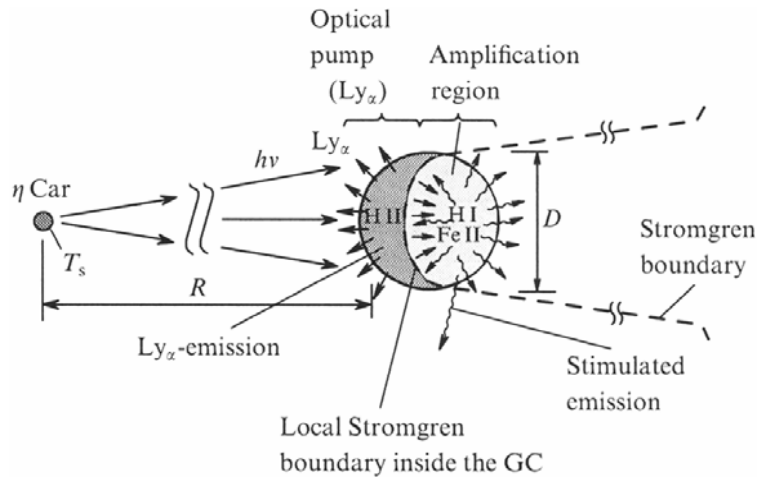

**Fig.6.** Outline of a compact gas condensation (GC) near the hot luminous star η Car with the Strömgren boundary between the photoionized (H II) and neutral (H I) regions. Intense Ly α radiation from H II regions is creating laser action through a pumping of Fe II atomic energy levels. The region of laser amplification is marked: Johansson & Letokhov [72]. Reprinted from JETP Lett **75**, 495, © 2002

Spectra of the bright condensations ("Weigelt blobs") in the η Car nebulosity display distinct Fe II emission lines, whose formation has been identified as due to stimulated emission: [73, 74]; Fig. 6. It is argued that laser amplification and stimulated emission must be fairly common for gaseous condensations in the vicinity of bright stars, caused by an interplay between fast radiative and slow collisional relaxation in these rarefied regions. These processes occur on highly different time scales, radiative relaxation operating over $10^{-9}$–$10^{-3}$ s, and collisional over seconds. In case of excitation of some high-lying electronic levels of an atom or ion with a complex energy-level structure, radiative relaxation follows as a consequence of spontaneous emission, in the course of which an inverse population develops of some pair(s) of levels, producing laser action.

Such emission is a diagnostic of non-equilibrium and spatially non-homogeneous physical conditions as well as a high brightness temperature of Ly α in ejecta from eruptive stars (Fig. 7).

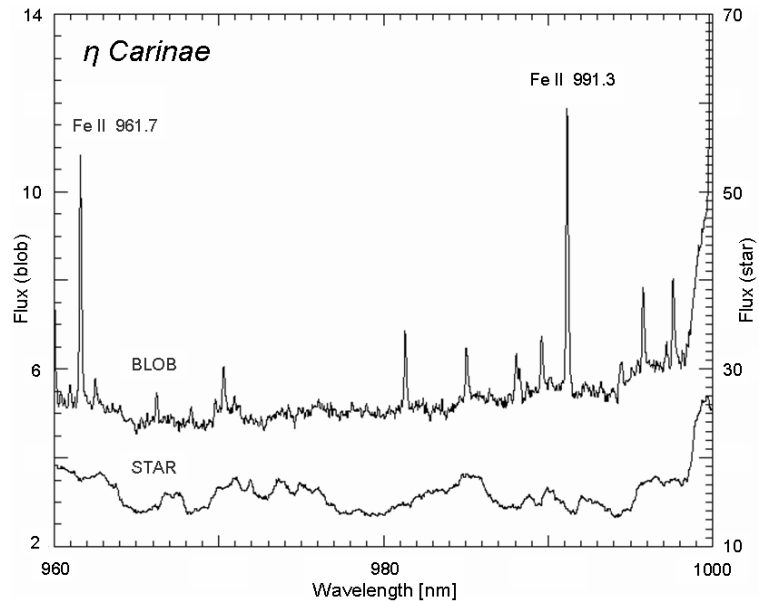

**Fig.7.** Spectra of the central star of η Car (flux scale at right) and the bright condensation in the nebulosity outside (Weigelt blob B; left scale), with the two lasing Fe II lines λ 961.7 and 991.3 nm: Johansson & Letokhov [74]. Reprinted from A&A **428**, 497 with permission from its publisher

A somewhat similar geometry is found in symbiotic stars, close binary systems where a hot star ionizes part of an extended envelope of a cooler companion, leading to complex radiative mechanisms. The combined spectrum shows the superposition of absorption and emission features together with irregular variability. In conditions where strong ultraviolet emission lines of highly ionized atoms (e.g., O VI λ 103.2, 103.8 nm) irradiate high-density regions of neutral hydrogen (with the Ly β line at λ 102.6 nm), Raman-scattered lines may be observed. Such have has been identified at λ 682.5 and 708.2 nm in the symbiotic star V1016 Cyg [135].

Stimulated emission does not necessarily require population inversions of atomic energy levels: Sorokin & Glownia [144] suggest "lasers without inversion" to explain the emission of a few narrow emission lines that dominate the visible or ultraviolet spectra of certain objects such as some symbiotic stars. Although the electronic level structures of the atoms/ions producing these emissions preclude the maintenance of population inversions, there are other ways to produce stimulated emission, where the atoms are made non-absorbing at the relevant lasing frequency ("electromagnetically induced transparency").

### 4.3 Hydrogen lasers in emission-line objects

MWC 349 is a peculiar hot emission-line star, often classified as B[e], with a B0 III companion. It is a very bright infrared source and an extremely strong radio star. The estimated mass is $\approx 30\ M_{Sun}$ and its distance $\approx 1.2$ kpc. It possesses a dense neutral Keplerian circumstellar disk (extent $\approx 300$ AU), and an ionized wind with very low terminal velocity ($\approx 50$ km/s). Proposed models include a quasi-spherical outflow of ionized gas in the outer part of the circumstellar envelope, and a differentially rotating disk viewed edge-on in its inner part [53].

Perhaps the most distinct features are strong hydrogen recombination line masers at mm and infrared wavelengths (H10α to H40α); [128]. The emission appears to be created by photoionization in a region with large electron density in a Keplerian disk at a distance ≈ 20−30 AU from the star [23]; Figs. 8−9.

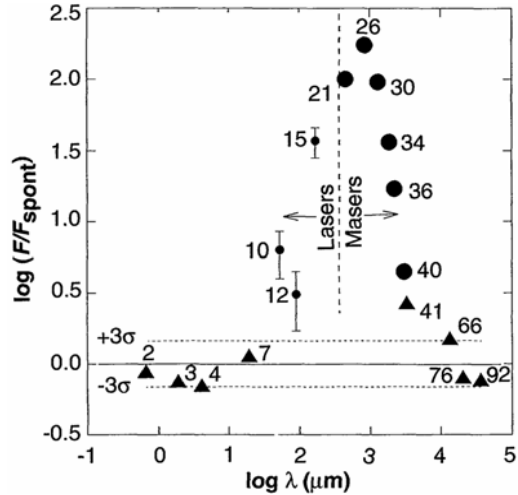

**Fig.8.** MWC349A: Log-log plot of the ratio $F/F_{spont}$ where F is the total observed flux in successive hydrogen recombination lines, and $F_{spont}$ the estimated contribution from spontaneous emission. Large dots indicate masing mm and sub-mm lines; small dots are infrared detections. The numbers are the principal quantum numbers for each line's lower level: Strelnitski et al [149]. Reprinted with permission from Science **272**, 1459, © 1996 AAAS

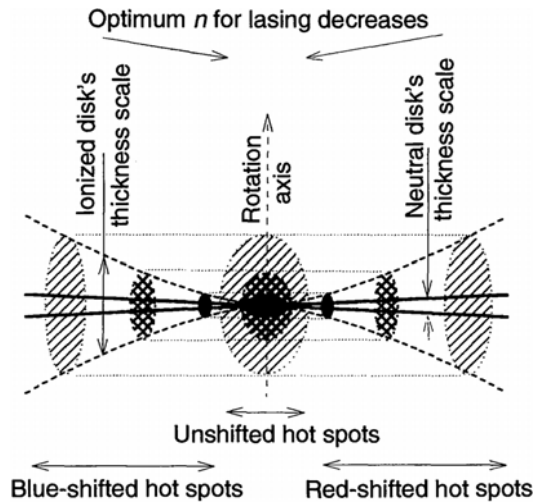

**Fig.9.** Proposed structure of the lasing and masing circumstellar disk in MWC349A. This edge-on view is presumably as we observe it: Strelnitski et al [149]. Reprinted with permission from Science **272**, 1459, © 1996 AAAS

### 4.4 Quantum optics in the earliest Universe

The physical processes exemplified by the lasers and masers in MWC349A may have been quite fundamental already in the *very* early Universe, soon after the Big Bang, at the epochs of recombination and reionization. The physical properties of masing hydrogen recombination lines at redshift $Z \approx 1000$ were modeled by Spaans & Norman [145].

Given small-scale overdense regions, maser action is possible owing to the expansion of the Universe and the low ambient temperature at radiative decoupling. Due to the redshift, these spectra would today be observable in the long-wavelength radio region (e.g., with the planned Square Kilometer Array, SKA), but the basic physical processes would be analogous to those observable at shorter wavelengths in today's local Universe.

### 4.5 $CO_2$ lasers in planetary atmospheres

Laser/maser emission is not a property that is exclusive to extreme deep-space environments, but can be found already in our planetary-system neighborhood (and even in the Earth's upper atmosphere!), in particular due to an interplay between solar ultraviolet radiation, and the energy levels in molecules such as $CO_2$.. Their observation ideally requires spectral resolutions in the range $10^6$ to $10^7$ [143].

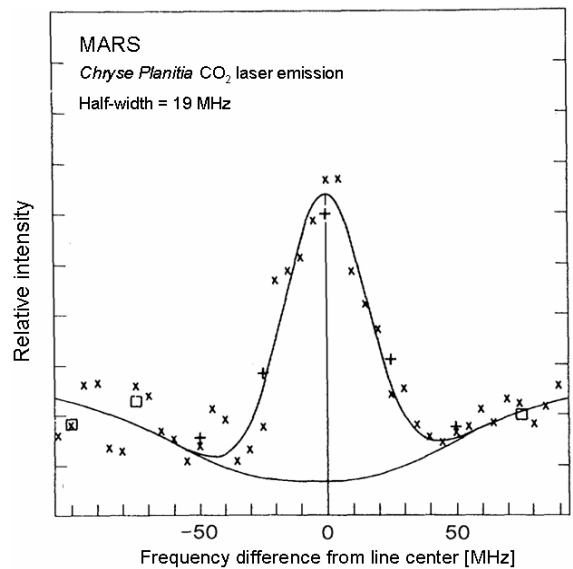

**Fig.10.** Spectra of a Martian $CO_2$ laser emission line ($^{12}C^{16}O_2$ R8 line, $\lambda$ 10.33 μm) as a function of frequency difference from line center (in MHz). Lower curve is the total emergent intensity in the absence of laser emission; the emission peak is modeled with a Gaussian fit: Mumma et al [107]. Reprinted with permission from Science **212**, 45, © 1981 AAAS

### 4.6 How short (and bright) pulses exist in nature?

The nanosecond structure observed within giant pulses from radio pulsars represents the currently most rapid fluctuations found in astronomical sources. Soglasnov et al [142] reported on observations of giant radio pulses from a millisecond pulsar at 1.65 GHz. Pulses were observed with widths ≤ 15 ns, corresponding to a brightness temperature of $T_b \geq 5\times10^{39}$ K, the highest so far observed in the Universe. Some 25 giant pulses are estimated to be generated during each neutron-star revolution. Their radiation energy density can exceed 300 times the plasma energy density at the surface of the neutron star and can even exceed the magnetic field energy density at that surface, constraining possible mechanism for their production.

Radiation mechanisms generating pulsar emission structure on very short timescales have been suggested already earlier [45]. If – as in classical textbooks – one would assume that the source size corresponds to the light-travel-time during the source variability, the structures responsible for pulses a few ns wide must be less than one meter in extent [63] (the speed of light is 30 cm ns$^{-1}$).

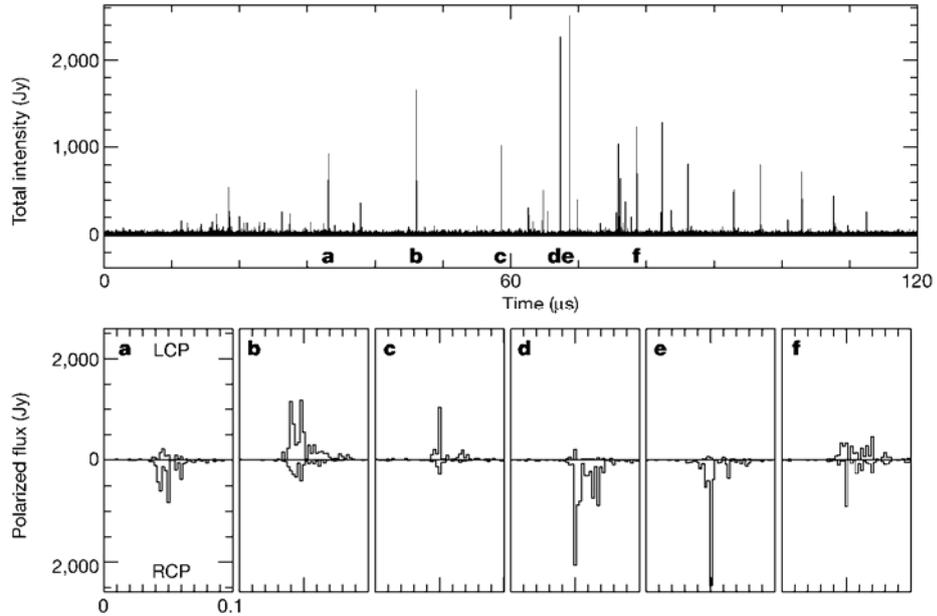

**Fig.11.** Nanosecond radio bursts from the Crab pulsar. Top: Details of a pulse seen with 2 ns resolution. Bottom: 100-ns sections, showing the left-, and right-circularly polarized flux from six of the nanopulses: Hankins et al [63]. Reprinted with permission from Macmillan Publishers Ltd: Nature **422**, 141, © 2003

Giant pulses have so far been observed only in some radio pulsars [36, 80, 82, 124], while theoretical ideas include energy release in nonlinear plasma turbulence [161], stimulated Compton scattering [118, 119], as well as effects from angular beaming arising due to relativistic motion of the radiating sources [46].

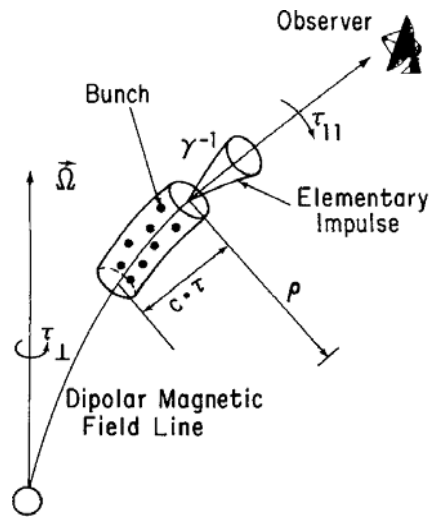

**Fig.12.** Possible mechanism for fluctuations on sub-microsecond timescales of pulsar emission, originating from relativistically moving particle bunches: Gil [45]. Reprinted from Ap&SS **110**, 293, with kind permission of Springer Science and Business Media, © 1985

Whether pulsars emit also *optical* flashes on nanosecond timescales is not clear. Searches for such pulses from the Crab pulsar were made already soon after its discovery, but without success. However, Shearer et al [139] detected a correlation between optical and giant radio pulse emission from the Crab. Optical pulses coincident with the giant radio pulses were on average 3% brighter than those coincident with normal radio pulses. This correlation suggests that there might also exist optical nanopulses, however, possibly requiring extremely large telescopes for their study.

## 4.7 Photon bubbles

When the intensity of light becomes very high, a volume filled with light takes on properties of a three-dimensional photon gas. Having a low mass density, it becomes buoyant, while exerting a pressure to its sides which may balance the surrounding material-gas pressure. In situations of very intense photon flux (e.g., in very hot stars) the development of *photohydrodynamic turbulence* involving the formation of "bubbles" of photon gas is expected. The bubbles would be filled with light, and would rise through the stellar surface, giving off photon bursts [66, 125, 138, 146]. Such phenomena are predicted to generate sub-millisecond fluctuations in accretion onto compact objects and can enable radiation fluxes orders of magnitude above the Eddington "limit" [7, 8, 79].

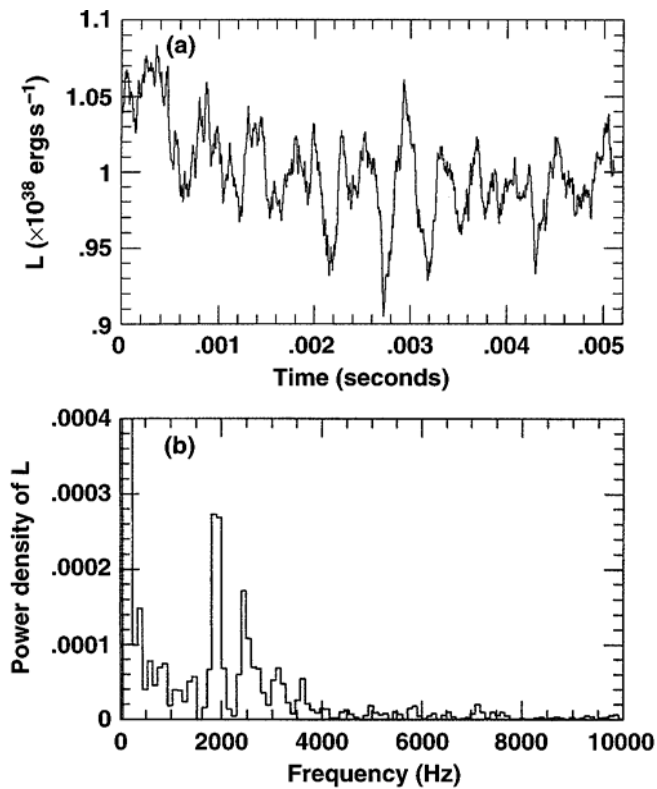

**Fig.13.** Theoretically simulated luminosity variations in a neutron-star accretion column due to photon-bubble oscillations: Klein et al [79]. Reproduced from ApJ **457**, L85 by permission of the AAS

## 4.8 Emission in magnetic fields of magnetars

Certain supernova explosions – those where the original star was spinning fast enough and had a strong enough magnetic field – can produce magnetars, highly magnetic neutron stars. Such objects may be most interesting targets for studying exotic radiation processes, including coherent laser effects. Normal pulsars can emit coherent radiation in the radio range, with a frequency corresponding to the plasma frequency in the pulsar magnetosphere; analogous processes in magnetar magnetospheres would have the plasma frequency appropriately scaled up, generating radiation in the optical or infrared.

A maser mechanism has been proposed to explain the optical pulsations from anomalous X-ray pulsars. Those would be generated by a curvature-drift-induced maser in the magnetar polar caps [99]. Another mechanism for coherent optical emission could be magnetospheric currents and plasma instabilities above magnetar surfaces [34]. Such coherent emission in the infrared and optical would probably be beamed, pulsed and polarized. Possibly, magnetospheric plasma could be produced in episodic sparking, on scales down to nanoseconds.

Magnetars are also among objects that may be suspected of hosting cosmic free-electron lasers (not involving inverted atomic energy-level populations but rather relativistic electrons in strong magnetic fields). Such lasers, having their own characteristic photon statistics, are a topic of wide current interest for various laboratory experiments [130, 136, 137].

## 5 Searching for Laser/Maser Effects

*Circumstantial* indicators for optical laser action include high line-to-continuum ratios; "anomalous" intensity ratios among lines in the same multiplet; rapid time variability (perhaps uncorrelated among different lines); small spatial extent, and narrow spectral lines (caused by the amplification of radiation at the line center being higher than in the wings), coupled with theoretical expectations.

Since any region radiating into a specific direction at any moment in time must be very much smaller than the full stellar atmosphere, there will exist a large number of independent emission regions, and Doppler shifts from stellar rotation or expanding atmospheres will smear out the integrated line profile, masking the spectrally narrow contributions from individual emission regions. Also, if only a small fraction of the source is lasing, spontaneous emission from the rest may overwhelm the small laser "hot-spots" when observed with insufficient spatial resolution. Still, suitable interferometry could reveal the existence of such hot-spots, even if they are continually changing [84, 85].

In radio, masers have been studied in great detail, and have become an important part of radio astrophysics for both local sources [35], and for the distant Universe [96]. Some examination has been made also of the statistics of radiation amplitudes. In radio, where photon-counting is not (yet) feasible, this corresponds to photon statistics in the optical. Ordinarily, however, the radiation is assumed to arise through processes in which individual particles radiate independently, and the resulting radiation fields then become chaotic (Gaussian). This assumption greatly simplifies the radiative transfer equations, although a few attempts to detach from this assumption have been made [165, 166]: in stimulated emission, the nonlinear gain characteristics may well disturb the Gaussian nature of the original field.

Early observational efforts to search for deviations from Gaussian statistics in coherent radio sources were made by Evans et al [37] who studied OH masers, concluding that the statistics were Gaussian within at least some percent. Likewise, Paschenko et al [115], and Lekht et al [89] found the statistics of OH sources to have noiselike signal properties. DeNoyer & Dodd [26] noted that a way to detect the non-Gaussian electromagnetic fields in saturated masers would be to perform statistical tests in individual frequency channels; some studies of OH masers, however, did not reveal any such signals.

A claimed detection of non-Gaussian statistics was reported in the radio emission from some pulsars by Jenet et al [71]. The large Arecibo telescope was used to measure the statistics from four pulsars using high time resolution (100 ns), detecting temporally coherent emission in three of these. However, Smits et al [141], in repeating similar measurements, concluded that the coherent signatures more likely originate from interstellar scintillation, rather than being intrinsic to the pulsars.

# 6 Modeling Astrophysical Photon Statistics

Analogous to past observational ventures into previously unexplored parameter domains, the search for phenomena of photon statistics in astronomical light is an explorative scheme where one does not know beforehand exactly what to expect, nor in which sources. Also – judging from past experience – any more detailed development of relevant astrophysical theory is unlikely to occur until after actual observations are available. Nevertheless, some authors have made efforts towards this direction, and somewhat analogous physical situations may illustrate what challenges to expect.

The theoretical problem of light scattering in a macroscopic turbulent medium such as air is reasonably well studied. In particular, the equations of transfer for $I^2$ and higher-order moments of intensity $I$ have been formulated and solved: e.g., [155]. A familiar result implies that stars twinkle more with increasing atmospheric turbulence. The value of $\langle I \rangle$, i.e. the total number of photons transmitted per unit time may well be constant, but $\langle I^2 \rangle$ increases with greater fluctuations in the medium. The quantum-optical problem of scattering light against atoms is somewhat related, except that the timescales involved are now those of the coherence time of light, not those of terrestrial windspeeds.

Theoretical treatments of astrophysical radiative transfer have so far almost exclusively treated the first-order quantities of intensity, spectrum and polarization, and not the transfer of $I^2$ and higher-order terms. There are a few notable exceptions, however, like the analytical solution of the higher-order moment equation relevant for radio scintillations in the interstellar medium [88, 91, 92] and attempts to formulate the quantum mechanical description of the transfer of radiation, including non-Markovian effects (i.e. such referring to more than one photon at a time) in a photon gas [100, 101]; the transfer equation for the density matrix of phase space cell occupation number states [112, 134], the need to introduce concepts from non-linear optics [166], and other relevant formulations [21, 147, 160]; further there must be many pertinent papers in laboratory quantum optics. However, as opposed to the laboratory case, the often complex nature of astronomical sources may not lend itself to simple such treatments, and there does not yet appear to exist any theoretical predictions for specific astronomical sources of any spectral line profiles of higher-order than one (i.e. the ordinary intensity versus wavelength).

It appears that the equations for radiative transfer must be written in a *microscopic* form, i.e. considering each radiation, excitation, and scattering process on a photon[s]-to-atom[s]-to-photon[s] level, rather than treating only statistically averaged light intensities, opacities, etc. While it could be argued that the many radiation sources in any astronomical object would tend to randomize the emerging radiation (erasing the signatures of non-equilibrium processes), the number of independent sources that contribute to the observed photon signal in any given short time interval is actually very limited. If 1000 photons per millisecond are collected, say, those can originate from at most 1000 spatially distinct locations, which during that particular millisecond happened to beam their radiation towards the Earth. Perhaps, such 1000 photons could originate from only 990 sources, say, in which case some insight in the source physics could be gained through measurements of photon statistics.

There exist other types of quantum phenomena known in the laboratory, which quite conceivably might be detected in photon statistics and correlations from also some astronomical sources. These include emission cascades which could show up in cross correlations between emission successively appearing in different spectral lines in the same deexcitation cascade, or various collective effects of light-emitting matter ("superradiance", i.e. several *atoms* emitting together).

## 7 Photon Orbital Angular Momentum

Photons have surprisingly many properties: One single photon arriving from a given direction, of any given wavelength, still can have hundreds of different states since it may carry different amounts of orbital angular momentum! The linear momentum of electromagnetic radiation is associated with radiation pressure, while the angular momentum is associated with the polarization of the optical beam. Only quite recently has it been more widely realized that radiation may in addition carry also *photon orbital angular momentum*, POAM. There were some pioneering laboratory experiments already in the 1930's [10], but only recently has it become possible to measure POAM for also individual photons [87]. The amount of POAM is characterized by an integer $\ell$ in units of the Planck constant $\hbar$, so that an absorber placed in the path of such a beam will, for each photon absorbed, acquire an angular momentum $\ell\hbar$. The integer $\ell$ gives the POAM states of the photon and also determines – in a quantum information context – how many bits of information that can be encoded in a single photon. Since, in the laboratory, photons can now be prepared with $\ell$ up to at least the order of 300, it implies that single photons may carry [at least] 8 bits of information, of considerable interest for quantum computing, and a main reason for the current interest in these phenomena [15].

The POAM phenomenon can be illustrated on a microscopic level: a small particle can be made to orbit around the light beam's axis while, at the same time, the beam's angular momentum (circular polarization) causes the particle to rotate on its own axis [114]. Some astronomical applications were already suggested by Harwit [65], although it is not verified whether there exist actual astronomical sources where a significant fraction of the photons carry such properties.

However, POAM manipulation at the telescope may permit high-contrast imaging inside an "optical vortex", with possible applications in exoplanet imaging [41, 152, 153]. This utilizes that the cross-sectional intensity pattern of all POAM beams has an annular character (with a dark spot in the center) which persists no matter how tightly the beam is focused. In a sense, such light thus acts as its own coronagraph, extinguishing the light from the central star, while letting through that from a nearby exoplanet.

The spatial characteristics of POAM make it sensitive to spatial disturbances such as atmospheric seeing [116], although it is not obvious whether that sensitivity could also be somehow exploited to extract information about spatial structure in the light source itself.

Analogous phenomena can be recorded also in the radio range, where suitably designed three-dimensional antenna configurations can detect the wavefront curvature of helical radio waves, offering observables beyond those conventionally measured [9, 17, 18].

## 8 Role of Large Optical Telescopes

High time resolution astrophysics, quantum optics and large telescopes are closely connected since the sensitivity to rapid variability and higher-order coherence increases *very rapidly* with telescope size.

Light-curves basically become useless for resolutions below microseconds where typical time intervals between successive photons may even be longer than the time resolution, and where timescales of variability are both irregular and unknown. Instead studies have to be of the statistics of the arriving photon stream, such as its correlations or power spectra. All such statistical functions depend on (at least) the second power of the measured source intensity. Figure 14 compares the observed signal (*I* ), its square and fourth powers, for telescopes of different size. The signal for classical quantities increases with the intensity *I* ; the signal in power spectra and second-order (two-photon) intensity correlation as $I^2$; while the signal for a four-photon correlation equals the probability that four photons are recorded in rapid succession, and thus increases with the fourth power of the intensity. This very steep dependence makes the largest telescopes *enormously more sensitive* for high-speed astrophysics and quantum optics.

| Telescope diameter | | Intensity <I> | Second-order correlation <I²> | Fourth-order photon statistics <I⁴> |
|---|---|---|---|---|
| 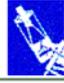 | 3.6 m | 1 | 1 | 1 |
| 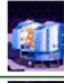 | 8.2 m | 5 | 27 | 720 |
| 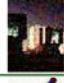 | 4 × 8.2 m | 21 | 430 | 185,000 |
| 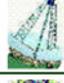 | 50 m | 193 | 37,000 | 1,385,000,000 |
| 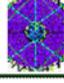 | 100 m | 770 | 595,000 | 355,000,000,000 |

**Fig.14.** Comparisons between the observed signal of source intensity *I*, its square and fourth powers, for telescopes of different size. The signal for classical quantities increases with the intensity *I*; the signal in power spectra and similar functions suitable for variability searches, as $I^2$, and that of four-photon correlations as $I^4$, as relevant for some quantum-statistical studies. The advent of very large telescopes enormously increases the potential for quantum optics, and high time resolution astrophysics in general

These large numbers may appear unusual when compared to the more modest gains expected for classical types of instruments, and initially perhaps even difficult to believe. Such numbers are, however, well understood among workers in non-linear optics. The measured $<I^4>$ is proportional to the conditional probability that four photons are recorded within a certain time interval. $<I^4>$ itself is, strictly speaking, not a physical observable: either one detects a photon in a time interval, or one does not. $<I^4>$ therefore has the meaning of successive intensity measurements in rapid succession: $<I(t)\cdot I(t+dt)\cdot I(t+2dt)\cdot I(t+3dt)>$. In an experiment where one is studying the multi-step ionization of some atomic species, where four successive photons have to be absorbed in rapid succession, one indeed sees how a doubling of the light intensity causes a 16-fold increase in the ionization efficiency. Or, for light of identical intensity, how the efficiency increases if the illuminating light source is changed to another of the same intensity but with different statistical properties, i.e. a different value of $<I^4>$.

But it does not stop there. The prospect of improved detectors will further increase the efficiency in a multiplicative manner. An increased quantum efficiency in the visual of a factor 3, say, or in the near infrared a factor 10, will mean factors of 10 and 100 in second-order quantities, while the signal in fourth-order functions will improve by factors 100 and 10,000, respectively. These factors should thus be *multiplied* with those already large numbers to give the likely gains for very large telescopes equipped with future photon-counting detectors, as compared to present ones.

Due to analogous steep dependences on intensity, the research field of non-linear optics was opened up by the advent of high-power laboratory lasers. In a similar vein, the advent of very large telescopes could well open up the field of very-high-speed astrophysical variability, and bring astronomical quantum optics above a detection threshold.

It is worth noting that – contrary to the situation in imaging – effects of atmospheric or telescopic *seeing* should not compromise measurements of intensity correlations. (This insensitivity to effects of atmospheric turbulence was indeed one of the main advantages of the original intensity interferometer.) To first order, atmospheric seeing induces only phase distortions and angular dislocations for the incoming light but does not affect the amount of illumination falling onto the telescope aperture. That, however, is modulated by the second-order effect of atmospheric *scintillation*. Characteristic timescales for scintillation fall in the range of milliseconds with only very little power remaining on scales of microseconds or shorter [28–30]. Nevertheless, any more precise search for very rapid astrophysical fluctuations will have to quantitatively check and calibrate the possible effects of the atmosphere, of any high-speed adaptive-optics systems used to compensate seeing, and of various transient atmospheric phenomena (such as meteors or Cherenkov radiation from energetic cosmic-ray particles).

## 9 Synergy with Large Radio Telescopes

*SKA*, the *Square Kilometer Array* is the large international radio telescope now being planned for longer radio wavelengths. One of its science drivers is strong-field tests of gravity using pulsars and black holes. The radio emission normally comes from lower-density regions in the magnetospheres around such objects, and expected studies of strong gravitational fields will focus on studying the orbit evolution of pulsars in orbit around other compact objects. The *SKA*'s sensitivity will likely enable the discovery of more than 10,000 pulsars in our Galaxy, including more than 1,000 millisecond ones. Many of these will probably be optically detectable with extremely large telescopes, and the *SKA* identifications will offer an excellent list of new targets for optical study.

Hydrogen recombination lasers/masers at the epoch of atomic recombination in the *very* early Universe ($Z \approx 1000$) were mentioned in Sect. 4.4 above. SKA might be able to observe these emission lines, strongly redshifted into the long-wavelength radio, even if those detections are likely to be difficult. On the other hand, optical large telescopes should permit studies of the equivalent physical processes of recombination lasers in nearby Galactic sources such as MWC 349A. This should place the interpretation of the cosmological emissions on a firmer physical basis, and could become an excellent example of the synergy between great telescopes operating in widely different wavelength regions.

Quantum effects are more readily reached at the longer radio wavelengths, where the number of photons per coherence volume can be very large. Indeed, the manifestation of the bunching of photons in the optical corresponds to "wave noise" in the radio. Such aspects also more readily produce conditions for maser emission in astronomical sources. Mechanisms such as the electron-cyclotron maser may operate in numerous classes of astronomical sources, from blazar jets near black holes, to exoplanets [154]. The very shortest nanosecond radio bursts seen from a few pulsars imply extremely high brightness temperatures and essentially reach the quantum limit (Fig. 15).

Thus, radio observations offer powerful tools that are complimentary to those in the optical. The optical probes higher-density regions close to stars and compact objects (rather than low-density clouds); optical spectral lines are those of atomic transitions (rather than those of low-temperature molecules), and the optical permits full studies of quantum phenomena that require the counting of individual photons, not yet practical at the longer radio wavelengths.

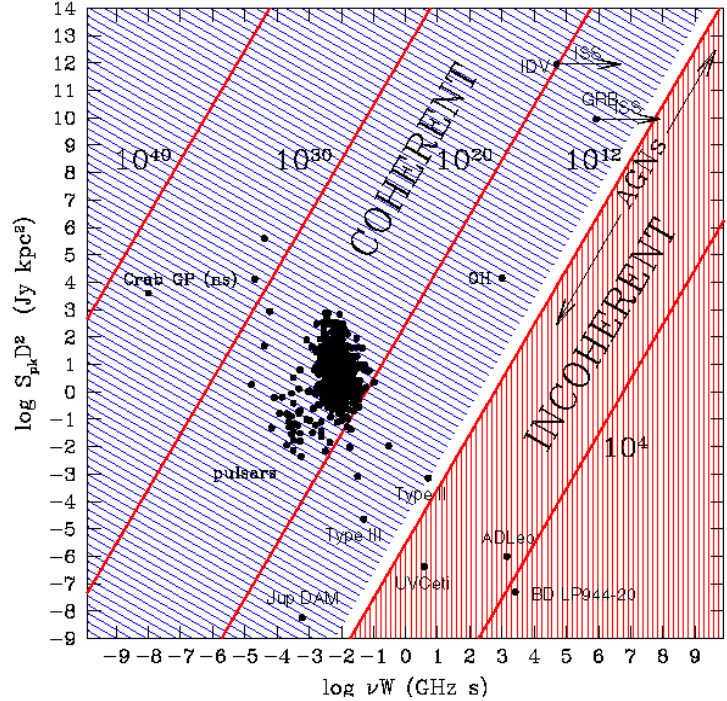

**Fig.15.** Phase space for known and anticipated transient radio signals. The horizontal axis is the product of transient duration $W$ and radio frequency $v$ while the vertical axis is the product of the observed peak flux density $S_{pk}$ and the square of the distance $D$, thus proportional to luminosity. Lines of constant brightness temperature [K] are shown. A quantum limit is set by the uncertainty principle which requires signals to have a duration of at least one cycle of electromagnetic radiation. With frequency $v$ given in units of $10^9$ Hz, this requires physical signals to be to the right of log $vW = -9$: Wilkinson et al, [164]. Reprinted from New Astron Rev **48**, 1551, © 2004, with permission from Elsevier

Radio observations permit various studies related to quantum statistics of radiation (some of which have already been pursued), e.g., the statistics of microwave maser radiation. Among future challenges remains the study of photon statistics of the cosmic microwave background; possibly this could be one future task in observational cosmology, once its currently sought polarization has been mapped?

## 10 Intensity Interferometry of Non-Photons

In an astronomical context, one naturally thinks of intensity interferometry in terms of correlations between streams of stellar photons. However, it may be illustrative to realize that in science and industry there exist widespread applications of this interferometric method pioneered by Hanbury Brown & Twiss, sometimes just referred to as "HBT-interferometry". The property being exploited, i.e., the "bunching" in time of photons applies not only to experiments in scattering of optical light [20] or of X-rays [167], but also to any other class of *bosons* (particles with integer spin), a phenomenon being applied in high-energy physics. The observed source is then not a star but perhaps the particles (e.g., pions, kaons) produced in an interaction region between high-energy particle beams. The angular size of the source can then be inferred from the cross correlation of intensity fluctuations observed by suitably placed detectors. The effort that has been invested in particle applications of intensity interferometry (hundreds of papers) is very much greater than that in the past Hanbury Brown-Twiss experiment for astronomy. Of course, in the laboratory, there are additional degrees of freedom, such as three-dimensional intensity interferometry of time-variable sources. For reviews, see Boal et al [11] and Alexander [2].

A somewhat different situation applies to fermions, i.e. particles with half-integral-valued spin, such as electrons. These obey Fermi-Dirac statistics and follow the Pauli exclusion principle which prohibits two or more particles to occupy the same quantum state. Thus, a gas of electrons or other fermions shows "anti-bunching", a property opposite to the bunching characteristic for thermal photons (Fig. 2 center). This is seen experimentally in the laboratory as a certain anti-correlation between nearby electrons [78].

It may be essential to appreciate these quantum properties of electrons when planning quantum-optical experiments. The many quantum-statistical properties, and different degrees of bunching in photon streams are permitted precisely by their boson nature and the more "permissive" Bose-Einstein statistics.

Although, using common language, one often speaks of photon "detection", *photons* as such are actually *never* directly detected! Rather, what is being detected and studied is some electrical signal from *photo-electrons* that results from some photon interaction inside the detector. It is a sobering thought that quantum-statistical properties to be measured, e.g., the bunching of several photons in the same quantum state, is a property that can not even in principle be possessed by these photo-electrons. One may then wonder how it is possible to at all study boson properties through a medium (electrons) that cannot, not even in principle, carry such properties? For photocathode detectors, the explanation apparently is the very short time (femtoseconds?) required for a photoelectron to exit a photocathode and then be detected as an individual particle.

However, semiconductor detectors are more complex, and have longer timescales for the relaxation of their inner energy levels [95]. A further discussion of possible ensuing complications is outside the scope of this paper (but see, e.g., [127]), though it is worth remembering that measurements of the quantum statistics of the incoming light may require an adequate understanding also of the quantum statistical properties of the detector through which these measurements are to be made: in a quantum world it may not always be possible to separate the observable from the observer.

## 11 Modern Intensity Interferometry

Astronomical observations with single or multiple large telescopes, involving the precise electronic timing of arriving photons within intense light fluxes enable various classes of intensity interferometry, and intensity correlation experiments. Given that contemporary electronic techniques are much more powerful than those originally used in the pioneering Narrabri instrument, one can envision various modern variants, and using only one single telescope aperture, spectroscopic measurements become feasible.

### 11.1 Photon-correlation spectroscopy

Photon-correlation (also called intensity-fluctuation, intensity-correlation or self-beating) spectroscopy is the temporal equivalent of spatial intensity interferometry. The cross-correlation of the optical fluctuations is then measured at one and the same spatial location, but with a variable temporal baseline (as opposed to the same instant of time but with a variable spatial baseline in the spatial intensity interferometer).

In any spectroscopic apparatus, the spectral resolution is ultimately limited by the Heisenberg uncertainty principle: $\Delta E \, \Delta t \geq \hbar/2$. Thus, to obtain a small value of the uncertainty in energy, $\Delta E$, the time to measure it, $\Delta t$, must be relatively great. Methods to increase $\Delta t$ include the use of larger diffraction gratings, and tilting them parallel to the direction of light propagation in order to increase the time light spends inside the instrument (this is the fundamental reason why echelle gratings give higher resolution than those in lower diffraction orders). Interference filters inside which light travels back and forth many times give high resolution (e.g., Fabry-Perot interferometers with a large finesse). Or – as in photon-correlation spectroscopy – instead of mechanical devices, one can use electronic timing of the light along its direction of propagation. Since this can be made for temporal delays up to perhaps one second, this enables a spectral resolution corresponding to that of a hardware instrument with physical size equal to one light-second! This makes possible spectral resolutions of 1 Hz, equivalent to R ≈ $10^{14}$, many orders of magnitude beyond those feasible with classical spectrometers (Table 1).

| *Spectral resolution R* | *Length* | *Time* |
|---|---|---|
| 100,000 | 6 cm | 200 ps |
| 1,000,000 | 60 cm | 2 ns |
| 10,000,000 | 6 m | 20 ns |
| 100,000,000 | 60 m | 200 ns |
| 1,000,000,000 | 600 m | 2 µs |

**Table 1.** Spectrometer length and equivalent light-travel-time requirements for different resolving powers at λ 600 nm

Somewhat paradoxically, the lower the spectral resolution, the more stringent the temporal requirements. The reason is that the timescale of intensity fluctuations is set by the self-beating time of the light within the detected passband. If this is 1 MHz, say, the characteristic time is its inverse (i.e., 1 µs), but for an astronomically more realistic optical passband of 1 nm, one is down to the order of ps, facing the same challenges as in spatial intensity interferometry. The method has become widespread for various laboratory applications to measure the very small Doppler broadenings caused by, e.g., exhaust fumes, molecular suspensions, or blood cells in living organisms undergoing random motions on a level of perhaps only mm s$^{-1}$.

Analogous to the spatial information extracted from intensity interferometry, photon-correlation spectroscopy does not reconstruct the full shape of the source spectrum, but "only" gives the power spectrum of the source with respect to the baseline over which it has been observed. In the case of the intensity interferometer, information is obtained on the relative power of spatial frequencies covered by the spatial baselines. In the case of the photon-correlation spectrometer, the result is the power of temporal [electromagnetic] frequencies covered by the temporal baselines, thus yielding the spectral linewidth but not automatically reconstructing the precise shape of a possibly asymmetric line profile. (Although the latter might actually be possible using more elaborate correlation schemes over multiple temporal baselines.)

The signal-to-noise ratios follow similar relations as in intensity interferometry, i.e., while generally being expensive in terms of photon flux, sources of high brightness temperature (such as narrow emission-line components) are easier to measure. Again similar to intensity interferometry, the method formally assumes Gaussian (chaotic; thermal; maximum-entropy) photon statistics, and will not work with purely coherent light. However, the analysis of superposed coherent and chaotic radiation could yield even more information [1, 157].

Technical discussions are in various papers of which several appeared around the time laser light scattering became established as a laboratory technique: [67–69, 81, 131–133]. For more general introductions to intensity-correlation spectroscopy principles and applications see Becker [6], Chu [20], Degiorgio & Lastovka [25], Oliver [113], Pike [122, 123], and Saleh [132].

Photon-correlation spectroscopy does not appear to have been used in astronomy, although there are analogs with [auto]correlation spectrometers used for radio astronomy.

While the basic physical principles are related, the latter build upon the detection of the phase and amplitude of the radio wave, an option not feasible for optical light due to its very much higher electromagnetic frequencies.

## 11.2 Spectroscopy with resolution R=100,000,000

Photon-correlation spectroscopy is especially suitable for the study of very narrow emission features. To resolve the infrared emission profiles from $CO_2$ lasers on Mars requires spectral resolving powers R = $\lambda/\Delta\lambda$ in the range of $10^6$–$10^7$ (Fig. 10), realized though various heterodyne techniques. The optical laser emission components around η Carinae are theoretically expected to have spectral widths $\Delta\nu$ on order 30–100 MHz [73, 75], thus requiring R ≈ $10^8$ (≈ 10 MHz) to be resolved, far beyond the means of classical spectroscopy. However, such emission lines will be self-beating on timescales on the order of $\Delta t$ ≈ 100 ns, well within reach of photon-correlation spectroscopy.

There might be a particular advantage for photon-correlation spectroscopy in that it is insensitive to (the probably rapidly variable) wavelength shifts of the emission line components due to local velocities in the source; one does not need to know exactly at which wavelength within the observed passband there might, at any one instant of time, exist some narrow emission-line components: they will all contribute to the correlation signal.

Possibly, the earliest laboratory measurement of a narrow emission line through photon-correlation spectroscopy was by Phillips et al [120] who, not long after the original experiments by Hanbury Brown & Twiss, deduced linewidths on the order of 100 MHz, corresponding to correlation times of some nanoseconds.

## 11.3 Optical *e*-interferometry

Using two or multiple telescopes (or telescope apertures), a precise electronic timing of arriving photons within intense light fluxes, combined with digital signal storage and handling, enables various new modes of electronic digital intensity interferometry, with degrees of freedom not available in the past. Discussions of such *e*-interferometry are in Dravins et al [31], who especially point at the potential of electronically combining multiple subapertures of extremely large telescopes, mainly for observations at short optical wavelengths. Pairs or groups of telescopes no longer need to be mechanically movable to track a star across the sky: the mechanical displacements can be replaced by continuously changing electronic delays. Ofir & Ribak [109–111] evaluate concepts for multidetector intensity interferometers. By digitally combining the signals across all possible pairs and triples of baselines (reminiscent of that in long baseline radio interferometry), not only is the noise level reduced and fainter sources accessible but, utilizing various proposed algorithms, one might achieve also the reconstruction of actual two-dimensional images and not merely their power spectra, as the case in classical two-element operations.

Since the required mechanical accuracy of the optics and stability of the atmosphere is on the order of centimeters (rather than a fraction of an optical wavelength, as in phase interferometry), long baselines may become feasible. Very long baseline optical interferometry (VLBOI; perhaps over 1–100 km) might thus be achieved by such electronic means; possible targets could include either nearby phenomena such as the granulation structure on the surfaces of nearby stars, or distant sources such as quasar cores. Initial experiments in VLBOI need not have dedicated telescopes but could be carried out with existing telescopes distributed across the grounds of adjacent observatories, or with flux collectors built primarily for measuring optical Cherenkov radiation in air from high-energy gamma rays, or perhaps even the coarser ones built for solar energy research.

Faster detectors are being developed, enabling higher sensitivity for such *e*-interferometry. However, faster electronics can only be exploited up to some point since there is a matching requirement on the optomechanical systems. A time resolution of 1 ns corresponds to 30 cm light travel distance, and the mechanical and optical paths then need to be controlled to only some reasonable fraction of this (perhaps 3 cm). However, a timing improvement to 100 ps, say, would require mechanical accuracies on mm levels, going beyond what is achieved in flux collectors, and beginning to approach the levels of seeing-induced fluctuations in atmospheric path-length differences.

Sources that are most advantageous to observe with intensity interferometry are those of small spatial extent with high brightness temperatures. These include suspected sites of stellar laser emission, where possible correlation studies were suggested already earlier [84, 85]. A recent paper [75] suggests to combine intensity interferometry with optical heterodyne spectroscopy (mixing the astronomical signal with a frequency-sweeping laboratory laser) to simultaneously identify spatial and spectral structures in the lasers around η Car.

## 12 Instrumentation for Extremely Large Telescopes

The practical detectability of higher-order effects in photon streams requires large photon fluxes as does indeed any study of very rapid variability. To identify nanosecond-scale variability will likely require sustained photon-count rates of many megahertz. This is one reason why the optical and infrared regions are most promising in searches for very rapid astrophysical variability. For many interesting sources, required count rates will be reached with foreseen optical telescopes, while the limited sizes of space instruments preclude such count rates for X-rays. While radio observations can be very sensitive, the absence of photon-counting could limit some classes of experiments; also the statistics of radiation can be strongly affected by the propagation through the ionized interstellar medium. Such considerations, as well as the dramatic increase in sensitivity to short-term variability (Fig. 14) indicate that high time resolution astrophysics and quantum optics are most suitable tasks for the coming generation of extremely large optical telescopes.

As part of the conceptual design studies of instruments for ESO's planned extremely large telescope, an instrument, *QuantEYE*, was conceived as the highest time-resolution instrument in optical astronomy [4, 31–33, 108], also described elsewhere in this volume [5]. *QuantEYE* is designed to explore astrophysical variability in the yet unexplored micro- and nanosecond domains, reaching the realm of quantum-optical phenomena. Foreseen targets include millisecond pulsars, variability close to black holes, surface convection on white dwarfs, non-radial oscillation spectra in neutron stars, fine structure on neutron-star surfaces, photon-gas bubbles in accretion flows, and possible free-electron lasers in the magnetic fields around magnetars. Measuring statistics of photon arrival times, *QuantEYE* is able to perform photon-correlation spectroscopy at resolutions $\lambda/\Delta\lambda > 100{,}000{,}000$, opening up for the first "extreme-resolution" optical spectroscopy in astrophysics, as well as to search for deviations from thermodynamic equilibrium in the quantum-optical photon bunching in time.

Irrespective of whether an instrument is eventually built following the precise concepts of *QuantEYE* (or whether it is overtaken by ongoing developments in detector technologies), its evaluation demonstrates how one can push the envelope of the parameter domains available to observational astronomy. During recent decades, these domains were mainly expanded through the opening up of successive new wavelength regions, each giving a new and different view of the Universe, thus supplementing the previous ones. However, now that almost all wavelength regions are accessible, the time domain becomes a bold new frontier for astrophysics, enabled by the forthcoming generation of extremely large optical telescopes, equipped with sensitive and fast photon-counting detectors, and supported by the now-established quantum theory of optical coherence!

# 13 Conclusions

During recent decades is has been realized, theoretically and in the laboratory that both individual photons, and groups of them, can be much more complex, and carry much more information than was commonly believed in the past. That even a single photon, of any given wavelength and polarization, and coming from any given direction, still can have hundreds of different states has come as a surprise to many who believed that photon properties were already well understood.

For astronomy, this poses both an opportunity and a challenge: since our understanding of the Universe is based upon the delicate decoding of information carried by light from celestial sources, we should exploit every opportunity to extract additional information, even if moving into uncharted and unknown territory, and not knowing beforehand what type of information will be conveyed. Quantum optics offers a new tool for studying extreme physical conditions in space, and the forthcoming extremely large optical telescopes promise to offer adequate sensitivity for exploiting quantum optics as another information channel from the Universe.

*Acknowledgements*

This work was supported by the Swedish Research Council, The Royal Physiographic Society in Lund, and the European Southern Observatory. During the design study of *QuantEYE*, numerous constructive discussions were held with colleagues at Lund Observatory, at the University of Padova, and at ESO Garching. Valuable comments on the manuscript were received from Henrik Hartman and Svenceric Johansson in Lund, Vladilen Letokhov (Lund & Moscow), and from an anonymous referee.